\documentclass[preprint]{elsarticle}
\usepackage[margin=2.8cm]{geometry}
\usepackage{graphicx, microtype, textcomp, booktabs, multirow}
\usepackage{amsmath,amssymb}
\usepackage[notrig]{physics}
\usepackage{isomath, mathtools}
\providecommand*{\eu}{\ensuremath{\mathrm{e}}}
\providecommand*{\iu}{\ensuremath{\mathrm{i}}}
\usepackage[range-units = single, range-phrase = \text{--}, separate-uncertainty = true, detect-all]{siunitx}
\usepackage[font=footnotesize]{subcaption}
\usepackage{hyperref}

\journal{JQSRT. \textcopyright 2017. This manuscript version is made available under \href{http://creativecommons.org/licenses/by-nc-nd/4.0/}{CC-BY-NC-ND 4.0} }

\begin{document}

\begin{frontmatter}

\title{CELES: CUDA-accelerated simulation of electromagnetic scattering by large ensembles of spheres}

\author[LTI,IMT]{Amos Egel\corref{cor1}}
\ead{amos.egel@kit.edu}
\author[FLO]{Lorenzo Pattelli}
\author[FLO,INO]{Giacomo Mazzamuto}
\author[FLO,UFI]{Diederik S. Wiersma}
\author[LTI,IMT]{Uli Lemmer}
\address[LTI]{Light Technology Institute, Karlsruhe Institute of Technology (KIT), Karlsruhe, Germany}
\address[IMT]{Institute for Microstructure Technology, Karlsruhe Institute of Technology (KIT), Karlsruhe, Germany}
\address[FLO]{European Laboratory for Non-linear Spectroscopy (LENS), 50019 Sesto Fiorentino, Florence, Italy}
\address[INO]{Istituto Nazionale di Ottica (CNR-INO), Via Carrara 1, 50019 Sesto Fiorentino, Florence, Italy}
\address[UFI]{Department of Physics, Universita di Firenze, 50019 Sesto Fiorentino, Florence, Italy}
\cortext[cor1]{Corresponding author}

\begin{abstract}
CELES is a freely available MATLAB toolbox to simulate light scattering by many spherical particles. Aiming at high computational performance, CELES leverages block-diagonal preconditioning, a lookup-table approach to evaluate costly functions and massively parallel execution on NVIDIA graphics processing units using the CUDA computing platform. The combination of these techniques allows to efficiently address large electrodynamic problems ($>$\num{e4} scatterers) on inexpensive consumer hardware.
In this paper, we validate near- and far-field distributions against the well-established multi-sphere $T$-matrix (MSTM) code and discuss the convergence behavior for ensembles of different sizes, including an exemplary system comprising \num{e5} particles.
\end{abstract}


\end{frontmatter}

\section{Introduction}
\label{sec:intro}
In computer-assisted investigations of light scattering and propagation, aggregates of spheres are traditionally used to represent various types of ordered and disordered optical materials. Systems that have been modeled as multi-sphere geometries include dust \cite{Xu1996} and soot particles \cite{Skorupski2013}, sand \cite{Born2015}, white paint \cite{Auger2003521}, photonic glasses \cite{GalisteoLopez2011, Romanov2016}, chiral structures \cite{AlJarro2016}, ice crystals \cite{Botta2013105}, arrays of plasmonic nano-particles \cite{Bakhti2016} and scattering layers in optoelectronic devices \cite{Egel2014a, Egel2016, Miranda-Munoz2016}.

Whereas the individual particles show a high degree of symmetry, structure is encoded in the relative particle configuration and size distribution. 
In the case of dilute particle ensembles, an individual-scattering approximation can be applied, which allows for a probabilistic ray optics description in combination with the Mie solution of single sphere scattering \cite{Mishchenko2011671, Voit2012, Mishchenko2013}. On the other hand, when particles are densely packed, coherent and near-field effects become important \cite{Sapienza2007, Okada2009902, Dlugach2011, RezvaniNaraghi2015, Gustavsson2016, Ma2017255, Ramezanpour2017} and a full wave-optics treatment of the multi-particle scattering problem is required. 

In this paper, we focus on the simulation of dense aggregates comprising large numbers of scattering particles. These simulations are usually employed to study bulk properties of scattering media, such as slabs or half-spaces of particles with a spatial dimension that is large compared to the extent of the probing beam.
When increasing the number of simulated particles, however, the computational load induced by multiple scattering grows rapidly. In order to push the limits of feasible ensemble sizes, one can either aim at more efficient algorithms, or at a better exploitation of available computer resources. Existing scattering codes for multiple spheres already offer parallel execution on computer clusters \cite{Mackowski2011}. With the release of a new code named CELES, we want to add a simulation environment that makes use of the massively parallel computing capabilities offered by consumer graphics processing units (GPUs). The purpose of this paper is to introduce the software, to demonstrate the correctness of the calculated fields, and to investigate the convergence behavior of simulations involving very large numbers of scattering particles. 

\section{Electromagnetic scattering by \texorpdfstring{$N$}{N} spheres}
\label{sec:theory}
The $T$-matrix formalism for the simulation of electromagnetic scattering by multiple particles has been described in many publications \cite{Fuller_AO_1991, Mackowski1996} (for spherical particles, this formalism is also referred to as the generalized multiparticle Mie-solution). Here, the theory is briefly summarized, mainly in order to establish the notation. We consider an ensemble of $N$ disjoint spheres $S_i$, each characterized by its center position $\vb*{r}_i$, its radius $R_i$ and complex refractive index $n_i$, $i=1\dots N$. The spheres are embedded inside a background medium with refractive index $n_0$. For simplicity, we assume that all materials are homogeneous, isotropic and non-magnetic. The particles are illuminated by a monochromatic incident field $\vb*{E}_\text{in}(\vb*{r})$ fulfilling Maxwell's equations in the absence of the scatterers. A harmonic time dependence $\exp(-\iu \omega t)$ is implicitly understood for all fields and we define the background wavenumber $k=n_0\omega/c$ with $c$ denoting the vacuum speed of light.

\subsection{Scattering by a single sphere}
In the case of electromagnetic scattering by one sphere, the $T$-matrix approach is equivalent to the well known Mie solution. Picking out one sphere $S_i$, we can write the total electric field as the sum of an incoming wave and the scattered field, which are expanded in terms of regular and outgoing spherical vector wave functions (SVWFs, see \ref{sec:VWFs}):

\begin{equation}
\vb*{E}(\vb*{r}) = \vb*{E}_{\text{in}}^i(\vb*{r}) + \vb*{E}_{\text{scat}}^i(\vb*{r})
\end{equation}
with
\begin{align}
\vb*{E}_{\text{in}}^i (\vb*{r}) &= \sum_{n}a_n^i \vb*{\Psi}_n^{(1)} (\vb*{r}-\vb*{r}_i) \\
\label{eq:Escat}
\vb*{E}_{\text{scat}}^i (\vb*{r}) &= \sum_{n}b_n^i \vb*{\Psi}_n^{(3)} (\vb*{r}-\vb*{r}_i) .
\end{align}
Here, $a^i_n$ and $b^i_n$ denote the SVWF coefficients of the incoming and the scattered field of the $i$-th sphere, respectively, while $n$ is a multi-index that subsumes the polarization $\tau=1,2$ and the multipole indices $l=1,2,\ldots$ and $m=-l,\ldots,l$. The $T$-matrix relates the coefficients of the incoming field to the coefficients of the scattered field:
\begin{equation}
b_n^i = \sum_{n'} T_{nn'}^i a_{n'}^i.
\label{eq:T-matrix}
\end{equation}
For isotropic spheres, $T^i_{nn'}$ is diagonal and does not depend on $m$. Explicit expressions are given in \ref{sec:mie}.

\subsection{Multiple scattering}
In the case of multiple particles, the incoming field for each particle $S_i$ is the sum of the initial excitation and the scattered field of all other spheres:
\begin{equation}
\vb*{E}_{\text{in}}^i (\vb*{r}) = \vb*{E}_{\text{in}} (\vb*{r}) + \sum_{i'\neq i} \vb*{E}_{\text{scat}}^{i'} (\vb*{r})
\end{equation}
Consequently, the incoming field coefficients are given by a contribution from the initial field plus a sum over contributions from all other particles. Whereas the former is known a priori (see \ref{sec:gaussianbeam} for a derivation of the initial field coefficients in the case of Gaussian beam illumination), the latter is a linear function of the scattered field coefficients of the other particles:
\begin{equation}
a^i_n = a_{\text{in},n}^i + \sum_{i'\neq i}\sum_{n'} W^{ii'}_{nn'} b_{n'}^{i'}.
\label{eq:a}
\end{equation}
Here, the coupling matrix $W$ is the transposed of the SVWF translation operator $A$ from $\vb*{r}_{i'}$ to $\vb*{r}_{i}$ (see \ref{sec:VWFs})
\begin{equation}
W^{ii'}_{nn'} = A_{n'n}(\vb*{r}_i-\vb*{r}_{i'}).
\end{equation}
Equations \eqref{eq:T-matrix} and \eqref{eq:a} form a coupled system of linear equations for $a_n^i$ and $b_n ^i$. Eliminating $a_n^i$ yields
\begin{equation}
\label{eq:masterequation}
\sum_{i',n'} M^{ii'}_{nn'} b_{n'}^{i'} = \sum_{n'} T^i_{nn'} a_{\text{in},n'}^i
\end{equation}
with
\begin{equation}
\label{eq:M}
M^{ii'}_{nn'} = \delta_{nn'}\delta_{ii'} - \sum_{n''} T^i_{nn''} W_{n''n'}^{ii'}.
\end{equation}
The multiple scattering problem is thereby reduced to the solution of the linear system of equations \eqref{eq:masterequation}. When the scattered field coefficients $b_n^i$ have been determined, all quantities of interest can be derived from them, including near and far-field distributions (see \ref{sec:far field}).

\section{The software}
\label{sec:software}
The CELES package is implemented in MATLAB, using an object oriented programming style. Code design was guided by the attempt to optimize the efficiency at the computational bottleneck (that is the solution of the linear system \eqref{eq:masterequation}) and following a ``keep it simple'' paradigm throughout the rest of the software design process.

The software is intended to simulate light scattering by large aggregates of spheres, where the ensemble of scattering targets is larger than the width of the incoming light ray. The appropriate initial excitation for the simulations is thus that of a Gaussian beam (although plane waves are implemented, too). Accordingly, the simulation output is given in terms of power reflectivity and transmittivity figures, as well as electric near field patterns and far field intensity distributions.

\subsection{Installation}
The CELES toolbox for the simulation of light scattering by many spherical particles is a free software distributed under the 3-Clause BSD License and can be downloaded from \url{http://github.com/disordered-photonics/celes}. In order to run simulations, the following system requirements need to be met:
\begin{itemize}
	\item A current MATLAB installation. The code was developed and tested using MATLAB 2016b.
	\item A CUDA-capable NVIDIA GPU.
	\item A CUDA toolkit installation consistent with the GPU model and MATLAB release. Use MATLAB's \texttt{gpuDevice} command to check for the compatible toolkit version.
	\item A C++ compiler that MATLAB accepts for CUDA compilation. Usually, on Linux platforms the built-in GCC C++ compiler is automatically detected and used. On Windows systems with MATLAB 2016b, the MS Visual Studio 2013 compiler needs to be installed.
\end{itemize}
If the system requirements are met, an exemplary simulation can be started by running the \texttt{CELES\_MAIN} script. Parameters that represent the particle configuration, the initial field as well as the numerical settings can be specified in that script following the instructions in the comments. 

\subsection{Computational strategy}
\label{ssec:strategy}
For very large numbers of particles, the matrix $M_{nn'}^{ii'}$ is too large to be stored in the main memory. Instead, we make use of the fact that for an iterative solution of the linear system \eqref{eq:masterequation}, only matrix-vector products are required. In the current version, the user can select between the biconjugate gradient stabilized method (BiCGSTAB) and the generalized minimal residual method (GMRES) \cite{Barrett1994}.
Then, the translation coefficients $A_{n'n}(\vb*{r}_i-\vb*{r}_{i'})$ can be computed on the fly during each iteration step, and do not need to be stored \cite{Mackowski2011}. Nonetheless, the convergence time of the iterative solver depends on the number of iterations needed to achieve some desired accuracy, and on the time that a single matrix-vector multiplication takes. Both factors grow with the number of considered particles. The computational strategy employed in the CELES software is thus based on three cornerstones to speed up the iterative solver: a block-diagonal preconditioner, a lookup table for the spherical Hankel function and GPU acceleration of the matrix vector-product evaluation.

\subsubsection{Block-diagonal preconditioner}
\begin{figure}[tb]
\centering
\includegraphics{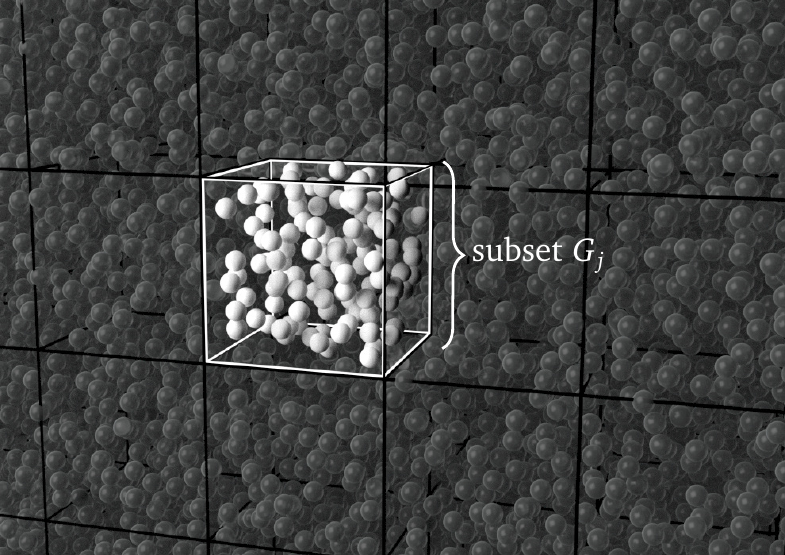} 
\caption{Graphical illustration of the block-diagonal preconditioner. The coupling between nearby particles is treated in terms of direct matrix inversion.}
\label{fig:preconditioner}
\end{figure}
The number of steps needed by an iterative solver to converge can be quite large, depending on the condition number of the linear system. One general strategy to improve the situation is to find a preconditioner, that is a map which approximates the inverse of the linear operator, and the computation of which takes much less time than the actual solution of the linear system itself. In order to construct a preconditioner, we take advantage of the fact that the strongest interaction occurs over short distances. The idea is thus to divide the sphere cluster into subgroups of neighboring particles and treat the interaction inside each of these groups in terms of a direct solution of the respective linear sub-system. 
In practice, this implies the following steps:
\begin{enumerate} 
	\item
	Dividing the set of spheres into $N_G$ subsets $G_j$, $j=1,\ldots,N_G$, see Figure \ref{fig:preconditioner}. Each subset contains $N_j$ spheres, such that $\sum_{j} N_j = N$. The subsets are constructed by dividing the volume occupied with spheres into an array of cuboids. For simplicity, the order of sphere indices $i$ is rearranged such that one subset $G_j$ corresponds to one successive series of sphere indices $i_{j}, \ldots, i_{j+N_j-1}$.
	\item 
	Computing the block matrices $M_j = M_{nn'}^{ii'}$ with $i_j \leq i,i' \leq i_{j+N_j-1}$.
	\item 
	Computing the $LU$-factorization for each block, $P_j M_j=L_j U_j$, where $P_j$ is a permutation matrix, and $L_j$ and $U_j$ are lower and upper triangular matrices, respectively. The matrices $P_j$, $L_j$ and $U_j$ are stored.
	\item
	The preconditioner is then a block-diagonal operator with blocks $M_j^{-1}$. In practice, the multiplication by a block, $x=M_j^{-1}y$ is evaluated by solving the system $L_j U_j x = P_j y$.
\end{enumerate}
Note, however, that the memory occupied by the storage of the $LU$ matrices scales as $\sum_j N_j^2$. This limits the possible size of the subsets $G_j$ when simulating very large particle aggregates. The size of the cuboids defining the subsets $N_G$ is provided by the user as an input parameter. We recommend the user to play with the partition edge size to find a reasonable trade-off between convergence rate and memory consumption.
\begin{figure*}
	\begin{subfigure}{.49\textwidth}
		\centering
		\includegraphics{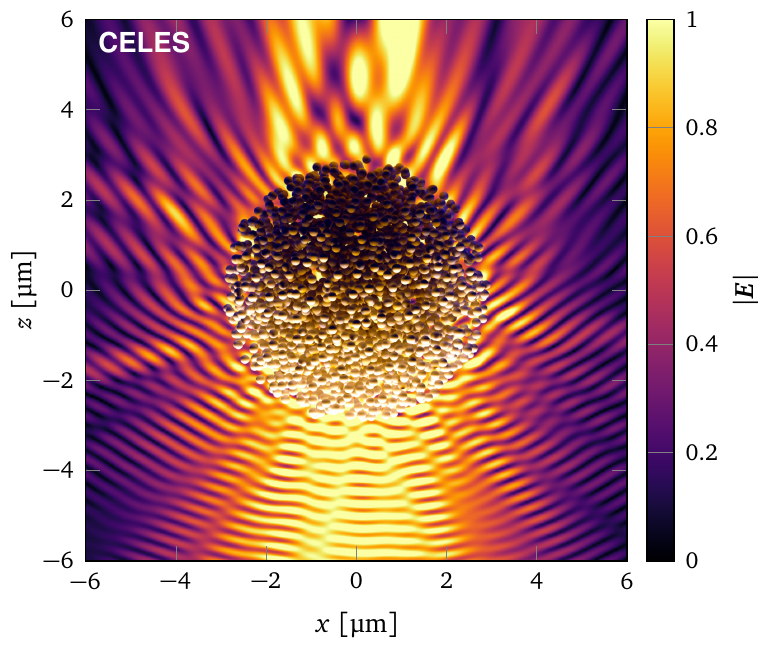}
		\label{fig:nf_celes}
	\end{subfigure}
	\begin{subfigure}{.49\textwidth}
		\centering
		\includegraphics{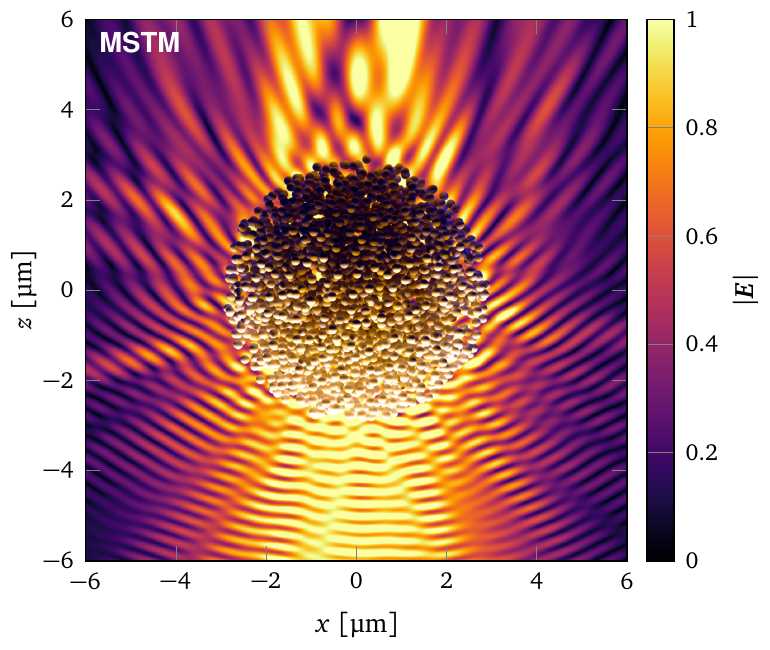}
		\label{fig:nf_mstm}
	\end{subfigure}
	\caption{Norm of the electric field, computed along the $y=\SI{0}{\micro\meter}$ plane with CELES (left) and MSTM (right) for a target sample made of \num{2.5e3} spheres. A 3D rendering of the spheres is superimposed to the field distribution to illustrate the configuration.}
	\label{fig:nf_validation}
\end{figure*}
\subsubsection{Matrix-vector product}
Now we turn to the actual evaluation of the matrix-vector products of type 
$\sum_{i',n'} M^{ii'}_{nn'} x_{n'}^{i'}$.
In the limit of large $N$, the computationally most intensive part is the product $\sum_{i',n'} W^{ii'}_{nn'} x_{n'}^{i'}$, with an effort scaling as $N^2$ (in contrast, the effort caused by the subsequent multiplication with the $T$-matrices scales linearly with $N$, as in \eqref{eq:M} $T^i_{nn'}$ does not depend on $i'$). It is thus sufficient to optimize the translation operator run time. The following methods are applied to achieve a good computational speed:
\begin{itemize}
	\item 
	We run the matrix-vector product $\sum_{i',n'} W^{ii'}_{nn'} x_{n'}^{i'}$ on the GPU by assigning one thread to each receiving particle $i$. The corresponding section of the code is implemented on NVIDIA's CUDA platform. The interface to the CUDA C kernel is provided by MATLAB's \texttt{mexcuda} environment. As a consequence of having one thread per particle, a good occupancy of the GPU is only achieved for high numbers of particles. Therefore, CELES runs most efficiently for large particle numbers. Because consumer graphic cards are in many cases optimized for single precision arithmetic, CELES is also implemented to run most operations in single precision in order to fully take advantage of the performance boost. As demonstrated in section \ref{ssec:validation}, the accuracy of the simulation results is not significantly affected.
	\item 
	The coefficients $a_5 (l,m|l',m'|p)$ and $b_5 (l,m|l',m'|p)$ in \eqref{eq:A operator} and \eqref{eq:B operator} involve the costly evaluation of square roots and Wigner-3j functions. However, as they do not depend on $i$ or $i'$, they are evaluated only once and stored in a table. Efficiency of this operation is therefore not critical.
	\item 
	The spherical Hankel function $h_p^{(1)}\qty(kd)$ depends only on the radial distance coordinate $d$. We precompute this function and store it in a table. On the GPU, cubic splines are used to interpolate the lookup table with good accuracy. The user can set the spatial resolution $\mathrm{\Delta} r$ of the lookup table. Whereas a very fine resolution has a slightly negative effect on the computational performance, a too coarse resolution can affect the accuracy. We recommend the user to play with this parameter in order to find a reasonable trade-off.
	\item
	The associated Legendre functions $P_p^{\abs{m-m'}} \qty(\cos\theta_d)$ are polynomials in $\cos\theta_d$ and $\sin\theta_d$ such that they can be quickly evaluated on the GPU. The coefficients of these polynomials are precomputed and stored in a table.
\end{itemize}

\section{Application examples}
\label{sec:application}
In the following, two case studies are presented to probe the validity and the convergence speed of the simulations. Afterwards, we present simulation results for light scattering by a large target comprising \num{e5} spheres. In each case, the investigated aggregates consist of spheres with radius $R_i=\SI{100}{\nano\meter}$ and refractive index $n_i=\num{1.5}$ in vacuum ($n_0=1$), and the excitation is provided by a linearly polarized Gaussian beam with a beam waist of \SI{4}{\micro\meter} and a vacuum wavelength of $\lambda=\SI{532}{\nano\meter}$ (size parameter of the spheres $2\pi R_i/\lambda\approx 1.18$). The truncation multipole degree was set to $l_\text{max}=3$, and the plane wave expansion of the incident field and the scattered field during the power flux evaluation of the CELES simulations was sampled with a polar and azimuthal angle resolution of $\mathrm{\Delta}\beta=\num{2.5e-4}\pi$ and $\mathrm{\Delta}\alpha=\num{2e-3} \pi$, respectively. The lookup table for the spherical Hankel functions was prepared using a spatial resolution of $\mathrm{\Delta}r=\SI{1}{\nano\meter}$.
For the solution of the linear system, we employed the GMRES solver with a relative tolerance of~$\num{e-4}$. All simulations were run on a Linux workstation computer with \SI{64}{GB} RAM and a Maxwell NVIDIA\textsuperscript{\textregistered} GTX Titan X graphic card (3072 single precision CUDA cores, \SI{12}{GB} GDDR5 memory). The code has also been tested on a Maxwell GeForce GTX 980 Ti card (2816 single precision CUDA cores, \SI{6}{GB} GDDR5 memory) with similar performances.

\subsection{Validation}
\label{ssec:validation}
In order to demonstrate the quantitative accuracy of the software, we have performed an exemplary simulation both with the CELES software package and with the MSTM software package \cite{Mackowski2011}. The example target consists of $N=2500$ spheres packed using the Lubachevsky-Stillinger algorithm to yield a final volume density of \SI{10}{\percent} inside a spherical region centered at $\vb*{r}=0$. The incident beam is focused at the center of the target.
 
\begin{figure}%
	\centering
	\includegraphics{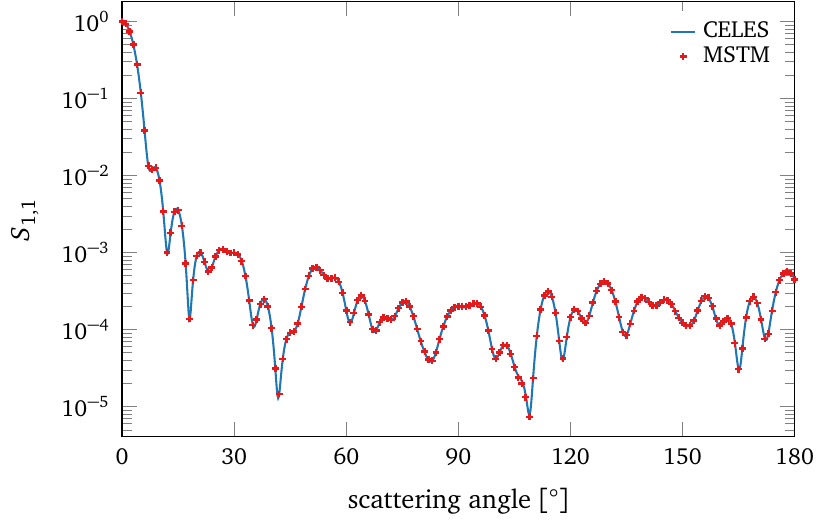}
	\caption{Far field intensity, computed with CELES and MSTM.}
	\label{fig:ff_validation}
\end{figure}

Figure \ref{fig:nf_validation} shows the norm of the resulting electric near-field distribution for the CELES and the MSTM simulations, exhibiting perfect agreement.
Figure \ref{fig:ff_validation} shows the $(1,1)$-element of the phase matrix, $S_{1,1}$, which was determined by running two simulations in CELES, one with an incoming TE-polarized beam and one with a TM-polarized beam, and then averaging the far field intensity distribution over both runs. On the other hand, MSTM returns the phase matrix directly (normalized to $S_{1,1}\qty(0)=1$). Also in this case the results are in excellent agreement. In addition, we have checked that the conservation of energy is fulfilled by verifying that the reflected power (\SI{16.25}{\percent}) plus the transmitted power (\SI{83.73}{\percent}) equals the incident power up to a relative error $<\num{e-3}$.

\subsection{Convergence behavior}
\begin{figure}[tb]%
\centering
\includegraphics{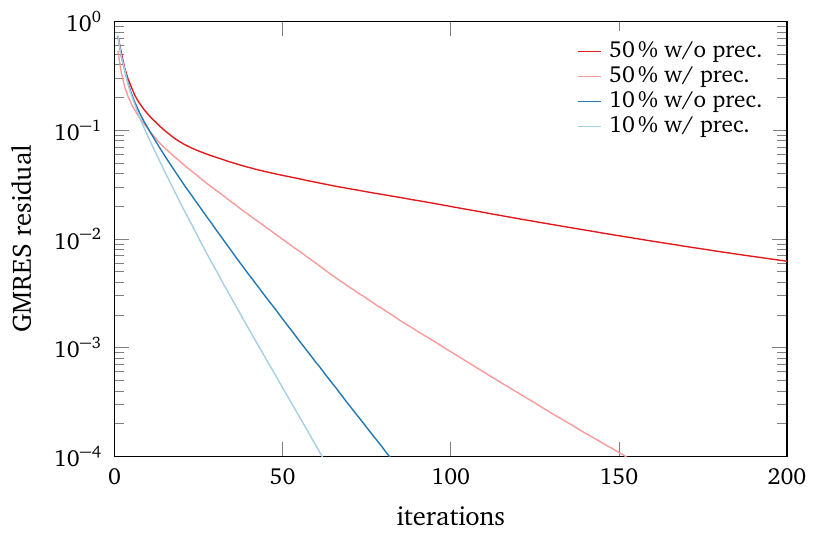}
\caption{Convergence of the GMRES iterative solver for aggregates of \num{2e4} particles at a volume fraction of \SI{10}{\percent} and \SI{50}{\percent}, each with and without preconditioner. The block-diagonal preconditioner is particularly effective for high-density aggregates.}
\label{fig:convergence}
\end{figure}%
\begin{figure}%
\centering
\includegraphics{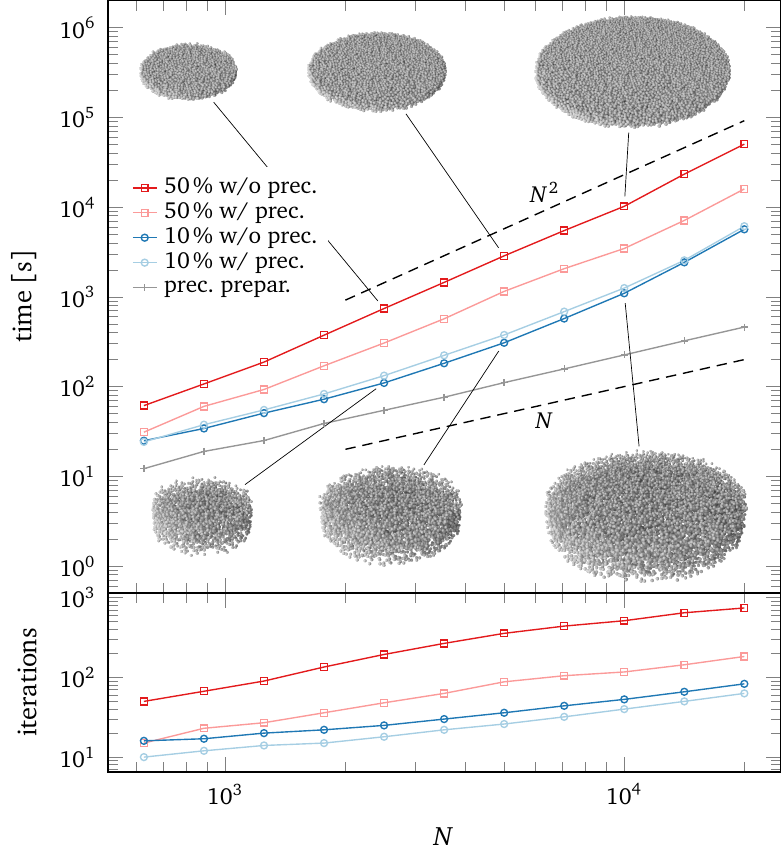}
\caption{Computation time and number of iterations versus system size. A few high and low-density configurations are shown for illustrative purposes.}
\label{fig:systemsize}
\end{figure}%
Another interesting aspect to discuss is the convergence speed and performance offered by CELES. 
Figure \ref{fig:convergence} shows the convergence of the solution for two slab configurations comprising \num{2e4} particles each, but with different overall volume fractions ($\text{vf} = \SI{10}{\percent}$ and $\text{vf} = \SI{50}{\percent}$). The initial Gaussian beam is focused on the outer surface of the cylindrical slab.
When active, the block-diagonal preconditioner is set so to divide the aggregate into cuboids containing roughly \num{200} particles each.

As can be seen, using the block-diagonal preconditioner offers limited advantage at lower volume fractions. Considering the $\text{vf} = \SI{10}{\percent}$ case, even though the number of iterations required is on average reduced by \SIrange{10}{20}{\percent}, the total run time is basically unchanged due to the overhead introduced by the preconditioner (cfr. Figure \ref{fig:systemsize}). The situation changes dramatically at higher densities ($\text{vf} = \SI{50}{\percent}$), where the convergence rate is much lower compared to the low-density samples. Then, using the preconditioner results in a significant reduction of the number of iterations and of the run time.
Figure \ref{fig:systemsize} shows how both these quantities grow with systems size.
The time needed to partition the system into several sub-groups and calculate a direct solution of each respective system grows linearly and becomes progressively inexpensive if compared to the overall simulation time for large aggregates.

It is interesting to compare the best run time obtained using CELES with that achievable by MSTM when leveraging all its speed-up techniques (i.e., far-field approximation and storing the translation matrix). For this comparison, we have used a workstation with the same amount of memory (\SI{64}{GB}) and 12 physical Xeon E5 2620 cores. We have checked that MSTM delivered best performance using a near-field translation distance of $kr=10$ and $\num{20}$ for the sparser and denser configurations, respectively.
With increasing particle number, the available memory became a limiting factor when using multithreading. For $\num{2e4}$ particles, we therefore needed to restrict MSTM to only 6 of the available 12 cores. The resulting runtimes are displayed in Table \ref{tab:runtimes}.
\begin{table}[h!]
\centering
\caption{\label{tab:runtimes}Run times comparison.}
\begin{tabular}{ccrr@{}c}
	\toprule 
	$N$						& density			& CELES 				& \multicolumn{2}{c}{MSTM (threads)}\\
	\midrule
	\multirow{2}*{\num{1e4}}& \SI{10}{\percent}	& \SI{1096}{\second}  	& \SI{2715}{\second} 	&(12) 		\\
							& \SI{50}{\percent}	& \SI{3463}{\second}  	& \SI{7574}{\second} 	&(12)		\\
	\multirow{2}*{\num{2e4}}& \SI{10}{\percent}	& \SI{5663}{\second}  	& \SI{29491}{\second}	&(6)		\\
							& \SI{50}{\percent}	& \SI{15951}{\second} 	& \SI{39380}{\second}	&(6)		\\
	\bottomrule
\end{tabular}
\end{table}

As a final note, in cases where convergence to a solution is particularly difficult to obtain, CELES offers the possibility to pass to the iterative solver a custom initial condition, typically represented by the solution of a smaller system comprising a sub-set of the total number of spheres.
\begin{figure*}
	\begin{subfigure}{.49\textwidth}
		\centering
		\includegraphics{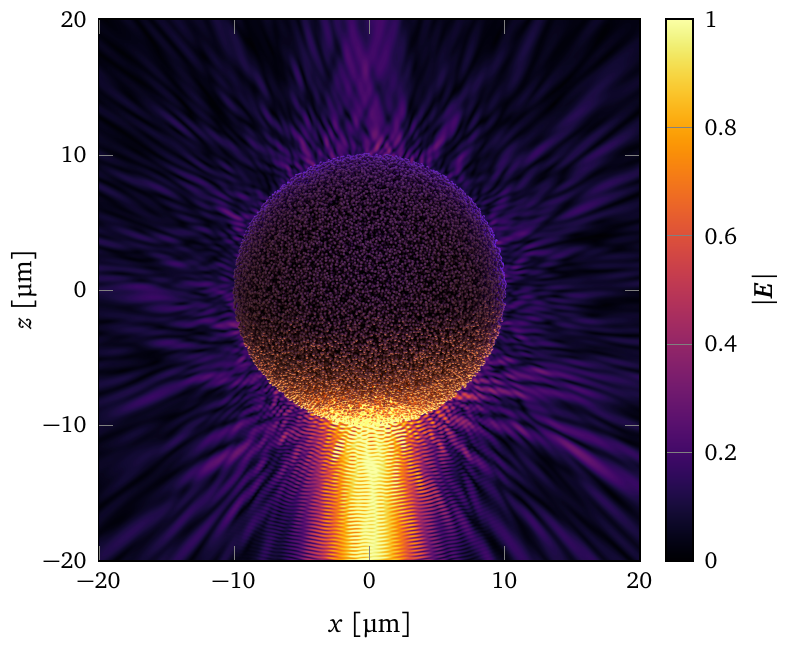}
	\end{subfigure}
	\begin{subfigure}{.49\textwidth}
		\centering
		\includegraphics{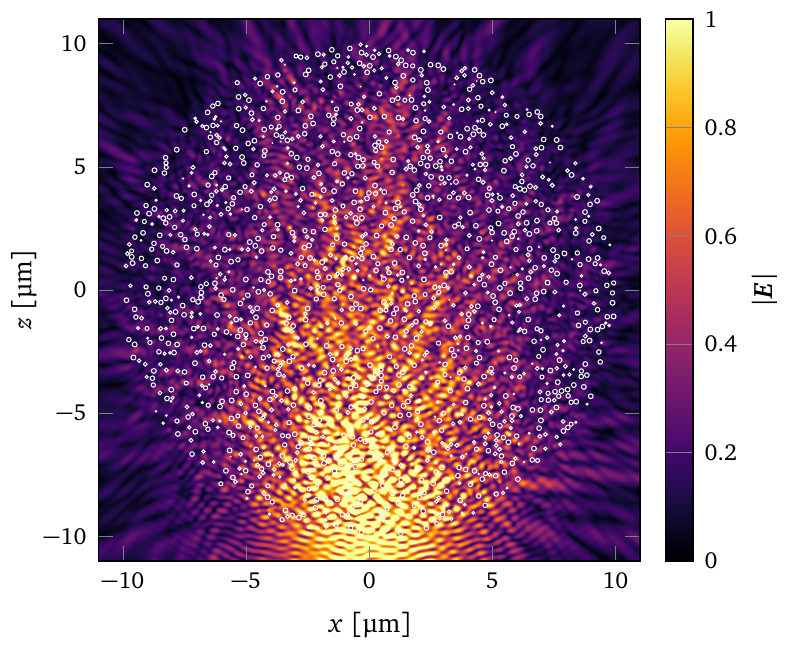}
	\end{subfigure}
	\caption{Norm of the electric field for a Gaussian beam scattered by \num{e5} spheres. In the left panel, a 3D rendering of the spheres is superimposed to the field distribution to illustrate the configuration. The right panel shows a magnification of the cross-cut plane $y=\SI{0}{\micro\meter}$ along which the field is calculated. White circles display the position of spheres cut by the image plane.}
	\label{fig:nf_100k}
\end{figure*}

\subsection{Large-scale target}
As we have seen in the last section, the effort to achieve convergence grows rapidly with the number of particles. However, we have successfully used CELES to run a simulation for a system comprising \num{e5} particles. Figure \ref{fig:nf_100k} shows the norm of the near field for a Gaussian beam hitting a spherical aggregate of \num{e5} particles at \SI{10}{\percent} volume density. The GMRES solver with block-diagonal preconditioner took \SI{3.15e5}{\second} ($\sim$\SI{87.5}{\hour}) to converge with a tolerance of \num{e-4}. Also in this case, the relative error with regard to energy conservation is less than \num{e-3}, with \SI{55.42}{\percent} of the incoming power scattered into the backward hemisphere and \SI{44.54}{\percent} transmitted or scattered into the forward hemisphere.

\section{Discussion}
\label{sec:discussion}
In the following, we will review and discuss some similarities and differences between CELES and existing codes or algorithms.

One of the most established software packages for multi-sphere scattering simulations is Mackowski's MSTM FORTRAN code \cite{Mackowski2011}, which is also freely available and which we used in our validation section. Supporting parallel execution on computer clusters, MSTM is also designed to allow for large particle numbers, and a far-field approximation can be switched on to accelerate the convergence of the iterative solver. A rotation-translation-rotation scheme is used for the SVWF translation, reducing the complexity of the matrix-vector product from $\mathcal{O} (l_\text{max}^4)$ to $\mathcal{O} (l_\text{max}^3)$ \cite{Mackowski1996}. In addition, MSTM is currently more flexible than CELES in that it allows for spheres inside other spheres and for chiral materials. It also offers the possibility to compute the ensemble $T$-matrix which in turn allows for an efficient orientation averaging. In comparison, some advantages of CELES are given by the unique speedup techniques described in section \ref{ssec:strategy}, which substantially enhance the calculation performance on a workstation computer equipped with a CUDA-capable GPU when addressing systems with a large number of particles.

Other implementations of electromagnetic multiple-sphere scattering are the FORTRAN GMM code by Xu and Gustafson \cite{Xu2001} (which has also been used to tackle scattering by $>$\num{e4} spherical particles \cite{Botta2013105}) and Pellegrini's py\_gmm package \cite{Pellegrini2007}, which offers python scripting capabilities and a user-friendly interface. 

Chew et al. \cite{chew1995fft} have proposed an aggregation of scattered field origins to a regular grid in combination with an acceleration of the matrix vector products based on the Fast-Fourier-Transform to achieve an effort that scales like $\mathcal{O} (N \log N)$.

Finally, the Fast Multipole Method (FMM) has been employed for the efficient simulation of wave scattering by large numbers of particles. It also brings a reduction of the complexity of matrix-vector products from $\mathcal{O} (N^2)$ to $\mathcal{O} (N \log N)$ \cite{Gumerov2005, gumerov2005fast}. Very good performances have been reported by Gimbutas and Greengard \cite{gimbutas2013fast}, as well as by Markkanen and Yuffa \cite{markkanen2017fast}. Both groups exploit FMM in combination with integral equation techniques to compute the individual particle $T$-matrices for large clusters of arbitrary-shaped particles.

With respect to CELES, it should be noted that the programmatic overhead introduced by FMM and other sophisticated techniques needs to be carefully considered, especially in the context of parallelization on GPU hardware. Occupancy and coalesced memory access are critical parameters for achieving a good computational performance and difficult to achieve when implementing more elaborate algorithms like FMM. Nevertheless, successful implementations of that kind have been demonstrated \cite{gumerov2008fast}. We thus believe that the combination of a Fast-Multipole scheme and GPU execution represents a promising direction to explore for future CELES releases.

\section{Conclusions}
The CELES software package is a new tool for the simulation of light scattering by large numbers of spherical particles. We have shown that 32 bit floating-point precision is sufficient to compute accurate near- and far-field distributions for large ensembles of scattering spheres, which opens up the possibility to exploit cost-effective non-scientific grade GPU hardware for this kind of calculations. An analysis of the convergence behaviour revealed that the application of a block-diagonal preconditioner is especially useful for the simulation of very dense particle aggreagates. As an open-source project, CELES is also open for contributions from other developers. Possible features for future releases include polydisperse particle samples, dipole source excitation or one of the advanced acceleration schemes for the matrix-vector product reviewed in section \ref{sec:discussion}.

\section*{Acknowledgements}
We wish to thank Daniel Mackowski for support during the validation process and Johannes Markkanen as well as Simone Zanotto for valuable hints and discussions. AE acknowledges financial support from the Karlsruhe House of Young Scientists (KHYS), the Karlsruhe School of Optics \& Photonics (KSOP) and from the DFG (SPP1839). DSW and LP acknowledge financial support from ERC Advanced Grant n.\ 291349.

\appendix

\section{Vector wave functions}
\label{sec:VWFs}
\subsection{Definition}
The spherical vector wave functions $\vb*{\Psi}^{(\nu)}_{\tau l m}$ live in the spherical coordinate system $(r,\theta,\phi)$ of the position vector $\vb*{r}$ and are defined as \cite{Doicu2006}
\begin{equation}
\begin{aligned}
\vb*{\Psi}_{1lm}^{(\nu)}(\vb*{r}) &= \frac{1}{\sqrt{2l(l+1)}} \curl{\left( \vb*{r} z_l^{(\nu)}(kr) P_l^{\abs{m}} (\cos\theta) \eu^{\iu m\phi}\right)} \\
\vb*{\Psi}_{2lm}^{(\nu)}(\vb*{r}) &= \frac{1}{k}\curl{\vb*{\Psi}_{1ml}^{(\nu)}(\vb*{r})}
\end{aligned}
\end{equation}
The number $(\nu)$ indicates if the SVWF is of regular kind $(\nu=1)$ or represents an outgoing wave $(\nu=3)$. Correspondingly, the radial wave function $z_l^{(\nu)}$ stands either for the spherical Bessel function of order $l$, $z_l^{(1)}=j_l$, or the spherical Hankel function of first kind, $z_l^{(3)}=h_l^{(1)}$. $P_l^m$ denote the normalized associated Legendre functions. Explicitly, the SVWFs read
\begin{equation}
\begin{aligned}
\vb*{\Psi}_{1lm}^{(\nu)}(\vb*{r}) = \frac{\exp(\iu m\phi)}{\sqrt{2l(l+1)}} & z_l^{(\nu)}(kr)\left( \iu m \pi_l^{\abs{m}}(\theta)\hat{\vb{e}}_\theta -\tau_l^{\abs{m}}(\theta)\hat{\vb{e}}_\phi \right) \\
\vb*{\Psi}_{2lm}^{(\nu)}(\vb*{r}) = \frac{\exp(\iu m\phi)}{\sqrt{2l(l+1)}} & \left(l(l+1)\frac{z_l^{(\nu)}(kr)}{kr}P_l^{\abs{m}}(\theta)\hat{\vb{e}}_r \right. \\
& \left. + \frac{\partial_{kr}\left( krz_l^{(\nu)}(kr)\right)}{kr}\left( \tau_l^{\abs{m}}(\theta)\hat{\vb{e}}_\theta + \iu m\pi_l^{\abs{m}}(\theta) \hat{\vb{e}}_\phi\right)\right),
\end{aligned}
\end{equation}
where
\begin{equation}
\begin{aligned}
\pi_l^m (\theta) &= \frac{P_l^m (\cos\theta)}{\sin\theta} \\
\tau_l^m (\theta) &= \partial_\theta P_l^m (\cos\theta).
\end{aligned}
\end{equation}
Further, the plane vector wave functions are defined as
\begin{equation}
\begin{aligned}
\vb*{\Phi}_1 (\alpha,\beta;\vb*{r}) &= \exp (\iu \vb*{k} \vdot \vb*{r}) \hat{\vb{e}}_\alpha \\
\vb*{\Phi}_2 (\alpha,\beta;\vb*{r})	&= \exp (\iu \vb*{k} \vdot \vb*{r}) \hat{\vb{e}}_\beta
\end{aligned}
\end{equation}
where $(k,\beta,\alpha)$ are the radial, polar and azimuthal spherical coordinate of the wavevector $\vb*{k}$, respectively.
\subsection{Translation and transformation}
The SVWF addition theorem accounts for the translation of the coordinate origin:
\begin{equation}
\vb*{\Psi}_n^{(3)}(\vb*{r}+\vb*{d}) = \sum_{n'} A_{nn'}(\vb*{d})\vb*{\Psi}_{n'}^{(1)}(\vb*{r}) \text{ for } \abs{\vb*{r}} < \abs{\vb*{d}}.
\end{equation}
The translation operator $A_{nn'}$ can be constructed iteratively \cite{Mackowski1991} or calculated from a closed form expression \cite{Stein1961, Cruzan1962, Bostrom1991, Mishchenko2002} involving the Wigner-3$j$ function:
\begin{equation}
A_{mlp,m'l'p'}(\vb*{d}) = \delta_{pp'} A_{ml,m'l'}(\vb*{d}) + (1-\delta_{pp'}) B_{ml,m'l'}(\vb*{d})
\end{equation}
with
\begin{align}
\label{eq:A operator}
A_{ml,m'l'}(\vb*{d}) &= \eu^{\iu (m-m')\phi_d} \sum_{\mathclap{p=\abs{l-l'}}}^{l+l'} a_5 (l,m|l',m'|p) h_p^{(1)} (kd) P_p^{\abs{m-m'}}(\cos\theta_d) \\
\label{eq:B operator}
B_{ml,m'l'}(\vb*{d}) &= \eu^{\iu (m-m')\phi_d} \sum_{\mathclap{p=\abs{l-l'}+1}}^{l+l'} b_5 (l,m|l',m'|p) h_p^{(1)} (kd) P_p^{\abs{m-m'}}(\cos\theta_d),
\end{align}
where
\begin{align}
\nonumber a_5 (l,m|l',m'|p) =& \iu^{\abs{m-m'}-\abs{m}-\abs{m'}+l'-l+p} (-1)^{m-m'} 
\\
&\times \left( l(l+1)+ l'(l'+1) - p(p+1)\right) \sqrt{2p+1} 
\\
&\times \sqrt{\frac{(2l+1)(2l'+1)}{2l(l+1)l'(l'+1)}} \Pmqty{l & l' & p \\ m & -m' & m'-m} \Pmqty{l & l' & p \\ 0 & 0 & 0}
\\
\nonumber b_5 (l,m|l',m'|p) =& \iu^{\abs{m-m'}-\abs{m}-\abs{m'}+l'-l+p} (-1)^{m-m'} 
\\
\nonumber &\times \sqrt{(l+l'+1+p)(l+l'+1-p)(p+l-l') (p-l+l')(2p+1)} 
\\
&\times \sqrt{\frac{(2l'+1)(2l+1)}{2l(l+1)l'(l'+1)}} \Pmqty{l & l' & p \\ m & -m' & m'-m} \Pmqty{l & l' & p-1 \\ 0 & 0 & 0}.
\end{align}
In the above, $(d,\theta_d,\phi_d)$ are the spherical coordinates of $\vb*{d}$, whereas $\sPmqty{\ldots \\ \ldots}$ denote the Wigner-3$j$ symbols.
In addition, the SVWFs can be transformed into PVWFs and vice versa. We make use of the following formulae:
\begin{align}
\label{eq:SVWF to PVWF}
\vb*{\Psi}_n^{(3)} (\vb*{r}) &= \frac{1}{2\pi} \int_{0}^{\mathrlap{2\pi}}\dd{\alpha} \int_{\mathrlap{C^\pm}} \dd{\beta} \sin\beta \sum_{j=1}^{2} B_{nj} (\beta) \vb*{\Phi}_j (\alpha,\beta;\vb*{r}) \eu^{\iu m\alpha} \\ \intertext{for $z \gtrless 0$, and}
\label{eq:PVWF to SVWF}
\vb*{\Phi}_j (\alpha,\beta;\vb*{r})	&= 4\sum_n \eu^{-\iu m\alpha} B_{nj}^\dagger
(\beta) \vb*{\Psi}_n^{(1)} (\vb*{r}). 
\end{align}
Here, the transformation operator $B_{nj}$ is given by
\begin{equation}\begin{aligned}
B_{nj}(\beta) = -\frac{1}{\iu^{l+1}}\frac{1}{\sqrt{2l(l+1)}} (\iu \delta_{j1}+\delta_{j2})\left( \delta_{\tau j}\tau_{l}^{\abs{m}}(\beta)+(1-\delta_{\tau j})m\pi_{l}^{\abs{m}}(\beta)\right),
\label{eq:B}
\end{aligned}\end{equation}
whereas $B_{nj}^\dagger$ has all explicit $\iu$ in \eqref{eq:B} set to $-\iu$. 

In \eqref{eq:SVWF to PVWF}, the contour $C^\pm$ of the $\beta$-integral is defined such that $\sin\beta$ runs from $0$ to $\infty$. In the case of $z>0$, $\beta$ starts at $0$ and goes to $\pi/2$ and then to $\pi/2-\iu \infty$ parallel to the imaginary axis, whereas in the case of $z<0$, $z$ starts at $\pi/2+\iu \infty$ and goes parallel to the imaginary axis to $\pi/2$ and then to $\pi$, compare \cite{Bostrom1991}.

\section{\texorpdfstring{$T$}{T}-matrix of a sphere}
\label{sec:mie}
The $T$-matrix of a sphere \cite{Bohren1983} is diagonal in all indices and its entries do not depend on the multipole order $m$.

\begin{align}
T_{nn'}^{i}= Q_{\tau l}^{i}\delta_{\tau\tau'}\delta_{mm'}\delta_{ll'}
\end{align}
with
\begin{align}
Q_{1l}^{i} &= \dfrac{j_{l} \left(k R_i\right) \partial_{k_i R_i} \left(k_i R_i j_{l} \left(k_i R_i \right) \right) - j_{l} \left(k_i R_i\right) \partial_{k R_i} \left(k R_ij_{l} \left(k R_i\right)\right)}{j_{l} \left(k_i R_i\right)\partial_{kR} \left(k R_ih_{l} \left(kR_i \right)\right)-h_{l}\left(kR_i \right) \partial_{k_iR_i} \left(k_iR_ij_{l} \left(k_iR_i \right) \right)}\\
Q_{2l}^{i}	& = \dfrac{ k^{2}j_{l}\left(kR_i\right) \partial_{k_iR_i} \left(k_i R_i j_{l} \left(k_iR_i\right)\right) - k_i^{2} j_{l} \left(k_iR_i\right) \partial_{k R_i}\left(kR_ij_{l}\left(kR_i\right)\right)} 
{k_i^{2}j_{l}\left(k_iR_i\right) \partial_{kR_i}\left(kR_ih_{l} \left(kR_i\right)\right) - k^{2}h_{l}\left(kR_i\right) \partial_{k_iR_i}\left(k_iR_ij_{l}\left(k_iR_i\right)\right)}.
\end{align}
In the above, $k_i=n_ik$ denotes the wavenumber inside the sphere.

\section{Gaussian beam}
\label{sec:gaussianbeam}
To mimic a Gaussian beam propagating into the positive $z$-direction, with a beam waist of width $w$, centered at $\vb*{r}_\text{G}=(x_\text{G},y_\text{G},z_\text{G})$, we require that for $z=z_\text{G}$ 
\begin{equation}\begin{aligned}
\vb*{E}_\text{G} (\vb*{r}) &= \exp \bqty{-\frac{(x-x_\text{G})^2+(y-y_\text{G})^2}{w^2}} \vb*{E}_0 \\
&= \int_{\mathbb{R}^2} \dd{k_x}\dd{k_y} \eu^{\iu (k_x x + k_y y)} \vb*{T}(k_x,k_y)
\label{eq:angular spec repr of Gauss}
\end{aligned}\end{equation}
where $\vb*{E}_0=(-\sin\alpha_G\hat{\vb{e}}_x+\cos\alpha_\text{G}\hat{\vb{e}}_y)E_0$ is a constant vector in the $xy$-plane and
\begin{equation}
\vb*{T}(k_x,k_y)=\frac{w^2}{4\pi}\exp\bqty{-(k_x^2+k_y^2)\frac{w^2}{4} -\iu (k_x x_\text{G}+k_y y_\text{G})}\vb*{E}_0,
\end{equation}
is the angular spectrum of the beam, compare \cite{Novotny2006}. Further, we use $k_x^2+k_y^2=k^2\sin^2\beta$ and $\dd{k_x}\dd{k_y} = k^2\dd{\alpha} \dd{\beta} \sin\beta \cos\beta$ and for $\beta \approx 0$ we use
\begin{equation}\begin{aligned}
\vb*{E}_0 &\approx \left(\cos(\alpha-\alpha_\text{G}) \hat{\vb{e}}_\alpha + 
\sin (\alpha-\alpha_\text{G}) \hat{\vb{e}}_\beta \right)E_0
\\
&= \left(\frac{\hat{\vb{e}}_\alpha -\iu \hat{\vb{e}}_\beta}{2}\eu^{\iu ( \alpha-\alpha_\text{G} )} +
\frac{\hat{\vb{e}}_\alpha + \iu \hat{\vb{e}}_\beta}{2}\eu^{-\iu ( \alpha-\alpha_\text{G} )} \right)E_0
\end{aligned}\end{equation}
to approximate \eqref{eq:angular spec repr of Gauss} by
\begin{equation}\begin{aligned}
\vb*{E}_\text{G} (\vb*{r}) &\approx \vb*{E}_\text{in} (\vb*{r}) \\
&= E_0 \frac{k^2w^2}{4\pi} \sum_{j=1}^2\int_0^{\mathrlap{\pi/2}} \dd{\beta} \sin\beta\cos\beta \exp\bqty{-\frac{w^2}{4}k^2\sin^2\beta}\\
&\hspace{0.5cm} \times \int_0^{\mathrlap{2\pi}} \dd{\alpha} \left( \eu^{\iu ( \alpha-\alpha_\text{G})} \frac{\delta_{j1}-\iu \delta_{j2}}{2} + \eu^{-\iu ( \alpha-\alpha_\text{G})} \frac{\delta_{j1}+\iu \delta_{j2}}{2}\right)  \eu^{\iu \vb*{k} \vdot (\vb*{r}_i-\vb*{r}_\text{G})} \vb*{\Phi}_j ( \alpha,\beta;\vb*{r}-\vb*{r}_i)
\label{eq:E_in Gauss}
\end{aligned}\end{equation}
We can use 
\begin{equation}
\eu^{ \iu \vb*{k}\vdot (\vb*{r}_i - \vb*{r}_\text{G})} = \exp\bqty{\iu k \rho_{\text{G},i}\sin \beta \cos (\alpha-\phi_{\text{G},i})} + \exp( \iu k z_{\text{G},i}\cos\beta)
\end{equation}
where $(\rho_{\text{G},i},\phi_{\text{G},i},z_{\text{G},i})$ are the cylindrical coordinates of $\vb*{r}_i-\vb*{r}_\text{G}$. 
Inserting \eqref{eq:PVWF to SVWF} into \eqref{eq:E_in Gauss} then yields
\begin{equation}
\begin{aligned}
\vb*{E}_\text{in}(\vb*{r}) = E_0 \frac{k^2w^2}{\pi} \sum_{j=1}^2 \sum_n \int_0^{\mathrlap{\pi/2}} \dd{\beta} & \sin\beta\cos\beta \exp \bqty{-\frac{w^2}{4}k^2\sin^2\beta } \eu^{\iu k\cos\beta z_{\text{G},i}} B^\dagger_{jn} (\beta) I_{jn}(\beta) \vb*{\Psi}^{(1)}_n (\vb*{r} - \vb*{r}_i)
\label{eq:E_in with Psi}
\end{aligned}
\end{equation}
where $I_j(\beta)$ denotes the $\alpha$-integral that can be evaluated analytically by using the identity $\int_0^{2\pi} \dd{\alpha} \eu^{\iu \nu\alpha} \eu^{\iu x\cos(\alpha-\phi)} = 2\pi\iu ^{\abs{\nu}} J_{\abs{\nu}} (x) \eu^{\iu \nu\phi}$ \cite{Stratton1941}.
\begin{equation}\begin{aligned}
I_{jn}(\beta) &= \int_0^{\mathrlap{2\pi}} \dd{\alpha} \left( \eu^{\iu(\alpha-\alpha_\text{G})} \frac{\delta_{j1} - \iu \delta_{j2}}{2} + \eu^{-\iu(\alpha-\alpha_\text{G})} \frac{\delta_{j1} + \iu \delta_{j2}}{2}\right) 
\eu^{-\iu m\alpha} \exp \bqty{\iu k \rho_{\text{G},i}\sin \beta \cos (\alpha-\phi_{\text{G},i})} \\
&= 2\pi \eu^{-\iu \alpha_\text{G}} \iu^{\abs{m-1}} \eu^{-\iu (m-1)\phi_{\text{G},i}} 
J_{\abs{m-1}} (k \rho_{\text{G},i}\sin \beta) \frac{\delta_{j1} - \iu\delta_{j2}}{2} \\
&\hspace{2.5cm}+ 2\pi \eu^{\iu \alpha_\text{G}} \iu^{\abs{m+1}} \eu^{-\iu(m+1)\phi_{\text{G},i}} J_{\abs{m+1}} (k \rho_{\text{G},i}\sin\beta) \frac{\delta_{j1} + \iu \delta_{j2}}{2}
\end{aligned}\end{equation}
Finally, by comparison of \eqref{eq:a} with \eqref{eq:E_in with Psi} one finds
\begin{equation}
a_{\text{in},n}^i = E_0 \frac{k^2w^2}{\pi} \sum_{j=1}^2\int_0^{\mathrlap{\pi/2}} \dd{\beta} \sin\beta \cos\beta B_{jn}^\dagger (\beta) I_j(\beta) \exp\bqty{-\frac{w^2}{4} k^2 \sin^2 \beta +\iu k z_{\text{G},i} \cos \beta} .
\end{equation}

\section{Far field intensity}
\label{sec:far field}
A field that is given by a plane wave expansion of the form
\begin{equation}
\vb*{E}(\vb*{r}) = \sum_{j=1}^2 \int \dd{\alpha} \int \dd{\beta} \sin(\beta) g_j (\alpha,\beta) \vb*{\Phi}_j(\vb*{r})
\end{equation}
gives rise to a radiant flux of \cite{Egel2014a}
\begin{equation}
\begin{aligned}
P &= \frac{2\pi^2}{\omega k \mu_0}\sum_{j=1}^2\int\dd{\alpha}\int\dd{\beta}\sin(\beta) \abs{g_j(\alpha,\beta)}^2 \\
&= \sum_{j=1}^2\int\dd{\alpha} \int\dd{\beta} \sin(\beta) I_j (\alpha,\beta),
\end{aligned}
\end{equation}
where
\begin{equation}
\label{eq:far field intensity}
I_j(\alpha,\beta)=\frac{2\pi^2 }{\omega k \mu_0} \abs{g_j(\alpha,\beta)}^2
\end{equation}
is the radiant intensity in the direction given by the polar angle $\beta$ and the azimuthal angle $\alpha$ with polarization $j$. 
The radiant intensity of the scattered field \eqref{eq:Escat} can thus be evaluated by first transforming it from the spherical wave expansion to a plane wave expansion using \eqref{eq:SVWF to PVWF} for each of the spheres, adding up the contribution of all spheres, and finally employing \eqref{eq:far field intensity}.	

Further, for a Gaussian beam \eqref{eq:E_in Gauss} we have
\begin{equation}
g_j(\alpha,\beta) = E_0\frac{k^2w^2}{4\pi}\cos\beta \exp\bqty{-\frac{w^2}{4}k^2\sin^2\beta}  (\delta_{j1}\cos\alpha + \delta_{j2}\sin\alpha) \eu^{-\iu\vb*{k}\vdot \vb*{r}_\text{G}}
\end{equation}
such that
\begin{equation}
P = \abs{E_0}^2 \frac{\pi k^3 w^4}{4 \omega \mu_0} \int_0^{\mathrlap{\pi/2}}\dd{\beta} \sin\beta \cos^2\beta \exp\bqty{-\frac{w^2}{2}k^2\sin^2\beta}.
\end{equation}

\section*{References}
\bibliography{celes_paper}

\begin{thebibliography}{43}
\expandafter\ifx\csname natexlab\endcsname\relax\def\natexlab#1{#1}\fi
\providecommand{\bibinfo}[2]{#2}
\ifx\xfnm\relax \def\xfnm[#1]{\unskip,\space#1}\fi
\bibitem[{{Xu} and {Gustafson}(1996)}]{Xu1996}
\bibinfo{author}{Y.-L. {Xu}}, \bibinfo{author}{B.~A.~S. {Gustafson}},
\newblock \bibinfo{title}{{A Complete and Efficient Multisphere Scattering
  Theory for Modeling the Optical Properties of Interplanetary Dust}},
\newblock in: \bibinfo{editor}{B.~A.~S. {Gustafson}}, \bibinfo{editor}{M.~S.
  {Hanner}} (Eds.), \bibinfo{booktitle}{{IAU Colloq. 150: Physics, Chemistry,
  and Dynamics of Interplanetary Dust}}, volume \bibinfo{volume}{104} of
  \textit{\bibinfo{series}{{Astronomical Society of the Pacific Conference
  Series}}}, p. \bibinfo{pages}{419}.
\bibitem[{Skorupski et~al.(2013)Skorupski, Mroczka, Riefler, Oltmann, Will, and
  Wriedt}]{Skorupski2013}
\bibinfo{author}{K.~Skorupski}, \bibinfo{author}{J.~Mroczka},
  \bibinfo{author}{N.~Riefler}, \bibinfo{author}{H.~Oltmann},
  \bibinfo{author}{S.~Will}, \bibinfo{author}{T.~Wriedt},
\newblock \bibinfo{title}{{Impact of morphological parameters onto simulated
  light scattering patterns}},
\newblock \bibinfo{journal}{J. Quant. Spectrosc. Radiat. Transf.}
  \bibinfo{volume}{119} (\bibinfo{year}{2013}) \bibinfo{pages}{53--66}.
\bibitem[{Born et~al.(2015)Born, Holldack, and Sperl}]{Born2015}
\bibinfo{author}{P.~Born}, \bibinfo{author}{K.~Holldack},
  \bibinfo{author}{M.~Sperl},
\newblock \bibinfo{title}{{Particle characterization using THz spectroscopy}},
\newblock \bibinfo{journal}{Granul. Matter} \bibinfo{volume}{17}
  (\bibinfo{year}{2015}) \bibinfo{pages}{531--536}.
\bibitem[{Auger et~al.(2003)Auger, Barrera, and Stout}]{Auger2003521}
\bibinfo{author}{J.-C. Auger}, \bibinfo{author}{R.~G. Barrera},
  \bibinfo{author}{B.~Stout},
\newblock \bibinfo{title}{{Scattering efficiency of clusters composed by
  aggregated spheres}},
\newblock \bibinfo{journal}{J. Quant. Spectrosc. Radiat. Transf.}
  \bibinfo{volume}{79--80} (\bibinfo{year}{2003}) \bibinfo{pages}{521--531}.
\bibitem[{Galisteo-L{\'o}pez et~al.(2011)Galisteo-L{\'o}pez, Ibisate, Sapienza,
  Froufe-P{\'e}rez, Blanco, and L{\'o}pez}]{GalisteoLopez2011}
\bibinfo{author}{J.~F. Galisteo-L{\'o}pez}, \bibinfo{author}{M.~Ibisate},
  \bibinfo{author}{R.~Sapienza}, \bibinfo{author}{L.~S. Froufe-P{\'e}rez},
  \bibinfo{author}{{\'A}.~Blanco}, \bibinfo{author}{C.~L{\'o}pez},
\newblock \bibinfo{title}{{Self-Assembled Photonic Structures}},
\newblock \bibinfo{journal}{Adv. Mater.} \bibinfo{volume}{23}
  (\bibinfo{year}{2011}) \bibinfo{pages}{30--69}.
\bibitem[{Romanov et~al.(2016)Romanov, Orlov, Ploss, Weiss, Vogel, and
  Peschel}]{Romanov2016}
\bibinfo{author}{S.~G. Romanov}, \bibinfo{author}{S.~Orlov},
  \bibinfo{author}{D.~Ploss}, \bibinfo{author}{C.~K. Weiss},
  \bibinfo{author}{N.~Vogel}, \bibinfo{author}{U.~Peschel},
\newblock \bibinfo{title}{{Engineered disorder and light propagation in a
  planar photonic glass}},
\newblock \bibinfo{journal}{Sci. Rep.} \bibinfo{volume}{6}
  (\bibinfo{year}{2016}) \bibinfo{pages}{27264}.
\bibitem[{Al-Jarro et~al.(2016)Al-Jarro, Biris, and Panoiu}]{AlJarro2016}
\bibinfo{author}{A.~Al-Jarro}, \bibinfo{author}{C.~G. Biris},
  \bibinfo{author}{N.~C. Panoiu},
\newblock \bibinfo{title}{Resonant mixing of optical orbital and spin angular
  momentum by using chiral silicon nanosphere clusters},
\newblock \bibinfo{journal}{Opt. Expr.} \bibinfo{volume}{24}
  (\bibinfo{year}{2016}) \bibinfo{pages}{6945--6958}.
\bibitem[{Botta et~al.(2013)Botta, Aydin, and Verlinde}]{Botta2013105}
\bibinfo{author}{G.~Botta}, \bibinfo{author}{K.~Aydin},
  \bibinfo{author}{J.~Verlinde},
\newblock \bibinfo{title}{Variability in millimeter wave scattering properties
  of dendritic ice crystals},
\newblock \bibinfo{journal}{J. Quant. Spectrosc. Radiat. Transf.}
  \bibinfo{volume}{131} (\bibinfo{year}{2013}) \bibinfo{pages}{105 -- 114}.
\bibitem[{Bakhti et~al.(2016)Bakhti, Tishchenko, Zambrana-Puyalto, Bonod,
  Dhuey, Schuck, Cabrini, Alayoglu, and Destouches}]{Bakhti2016}
\bibinfo{author}{S.~Bakhti}, \bibinfo{author}{A.~V. Tishchenko},
  \bibinfo{author}{X.~Zambrana-Puyalto}, \bibinfo{author}{N.~Bonod},
  \bibinfo{author}{S.~D. Dhuey}, \bibinfo{author}{P.~J. Schuck},
  \bibinfo{author}{S.~Cabrini}, \bibinfo{author}{S.~Alayoglu},
  \bibinfo{author}{N.~Destouches},
\newblock \bibinfo{title}{{Fano-like resonance emerging from magnetic and
  electric plasmon mode coupling in small arrays of gold particles.}},
\newblock \bibinfo{journal}{Sci. Rep.} \bibinfo{volume}{6}
  (\bibinfo{year}{2016}) \bibinfo{pages}{32061}.
\bibitem[{Egel and Lemmer(2014)}]{Egel2014a}
\bibinfo{author}{A.~Egel}, \bibinfo{author}{U.~Lemmer},
\newblock \bibinfo{title}{{Dipole emission in stratified media with multiple
  spherical scatterers: Enhanced outcoupling from OLEDs}},
\newblock \bibinfo{journal}{J. Quant. Spectrosc. Radiat. Transf.}
  \bibinfo{volume}{148} (\bibinfo{year}{2014}) \bibinfo{pages}{165--176}.
\bibitem[{Egel et~al.(2016)Egel, Gomard, Kettlitz, and Lemmer}]{Egel2016}
\bibinfo{author}{A.~Egel}, \bibinfo{author}{G.~Gomard},
  \bibinfo{author}{S.~Kettlitz}, \bibinfo{author}{U.~Lemmer},
\newblock \bibinfo{title}{{Accurate optical simulation of nano-particle based
  internal scattering layers for light outcoupling from organic light emitting
  diodes}},
\newblock \bibinfo{journal}{J. Opt.}  (\bibinfo{year}{2016}).
\bibitem[{Miranda-Mu{\~{n}}oz et~al.(2016)Miranda-Mu{\~{n}}oz,
  Carretero-Palacios, Jim{\'{e}}nez-Solano, Li, Lozano, and
  M{\'{i}}guez}]{Miranda-Munoz2016}
\bibinfo{author}{J.~M. Miranda-Mu{\~{n}}oz},
  \bibinfo{author}{S.~Carretero-Palacios},
  \bibinfo{author}{A.~Jim{\'{e}}nez-Solano}, \bibinfo{author}{Y.~Li},
  \bibinfo{author}{G.~Lozano}, \bibinfo{author}{H.~M{\'{i}}guez},
\newblock \bibinfo{title}{{Efficient bifacial dye-sensitized solar cells
  through disorder by design}},
\newblock \bibinfo{journal}{J. Mater. Chem. A} \bibinfo{volume}{4}
  (\bibinfo{year}{2016}) \bibinfo{pages}{1953--1961}.
\bibitem[{Mishchenko et~al.(2011)Mishchenko, Tishkovets, Travis, Cairns,
  Dlugach, Liu, Rosenbush, and Kiselev}]{Mishchenko2011671}
\bibinfo{author}{M.~I. Mishchenko}, \bibinfo{author}{V.~P. Tishkovets},
  \bibinfo{author}{L.~D. Travis}, \bibinfo{author}{B.~Cairns},
  \bibinfo{author}{J.~M. Dlugach}, \bibinfo{author}{L.~Liu},
  \bibinfo{author}{V.~K. Rosenbush}, \bibinfo{author}{N.~N. Kiselev},
\newblock \bibinfo{title}{{Electromagnetic scattering by a morphologically
  complex object: Fundamental concepts and common misconceptions}},
\newblock \bibinfo{journal}{J. Quant. Spectrosc. Radiat. Transf.}
  \bibinfo{volume}{112} (\bibinfo{year}{2011}) \bibinfo{pages}{671--692}.
\bibitem[{Voit et~al.(2012)Voit, Hohmann, Sch{\"a}fer, and Kienle}]{Voit2012}
\bibinfo{author}{F.~Voit}, \bibinfo{author}{A.~Hohmann},
  \bibinfo{author}{J.~Sch{\"a}fer}, \bibinfo{author}{A.~Kienle},
\newblock \bibinfo{title}{{Multiple scattering of polarized light: comparison
  of Maxwell theory and radiative transfer theory}},
\newblock \bibinfo{journal}{J. Biomed. Opt.} \bibinfo{volume}{17}
  (\bibinfo{year}{2012}) \bibinfo{pages}{045003--1--045003--8}.
\bibitem[{Mishchenko et~al.(2013)Mishchenko, Goldstein, Chowdhary, and
  Lompado}]{Mishchenko2013}
\bibinfo{author}{M.~I. Mishchenko}, \bibinfo{author}{D.~H. Goldstein},
  \bibinfo{author}{J.~Chowdhary}, \bibinfo{author}{A.~Lompado},
\newblock \bibinfo{title}{{Radiative transfer theory verified by controlled
  laboratory experiments}},
\newblock \bibinfo{journal}{Opt. Lett.} \bibinfo{volume}{38}
  (\bibinfo{year}{2013}) \bibinfo{pages}{3522--3525}.
\bibitem[{Sapienza et~al.(2007)Sapienza, Garc{\'i}a, Bertolotti, Mart{\'i}n,
  Blanco, Vi{\~n}a, L{\'o}pez, and Wiersma}]{Sapienza2007}
\bibinfo{author}{R.~Sapienza}, \bibinfo{author}{P.~D. Garc{\'i}a},
  \bibinfo{author}{J.~Bertolotti}, \bibinfo{author}{M.~D. Mart{\'i}n},
  \bibinfo{author}{{\'A}.~Blanco}, \bibinfo{author}{L.~Vi{\~n}a},
  \bibinfo{author}{C.~L{\'o}pez}, \bibinfo{author}{D.~S. Wiersma},
\newblock \bibinfo{title}{{Observation of Resonant Behavior in the Energy
  Velocity of Diffused Light}},
\newblock \bibinfo{journal}{Phys. Rev. Lett.} \bibinfo{volume}{99}
  (\bibinfo{year}{2007}) \bibinfo{pages}{233902}.
\bibitem[{Okada and Kokhanovsky(2009)}]{Okada2009902}
\bibinfo{author}{Y.~Okada}, \bibinfo{author}{A.~Kokhanovsky},
\newblock \bibinfo{title}{{Light scattering and absorption by densely packed
  groups of spherical particles}},
\newblock \bibinfo{journal}{J. Quant. Spectrosc. Radiat. Transf.}
  \bibinfo{volume}{110} (\bibinfo{year}{2009}) \bibinfo{pages}{902--917}.
\bibitem[{Dlugach et~al.(2011)Dlugach, Mishchenko, Liu, and
  Mackowski}]{Dlugach2011}
\bibinfo{author}{J.~M. Dlugach}, \bibinfo{author}{M.~I. Mishchenko},
  \bibinfo{author}{L.~Liu}, \bibinfo{author}{D.~W. Mackowski},
\newblock \bibinfo{title}{{Numerically exact computer simulations of light
  scattering by densely packed, random particulate media}},
\newblock \bibinfo{journal}{J. Quant. Spectrosc. Radiat. Transf.}
  \bibinfo{volume}{112} (\bibinfo{year}{2011}) \bibinfo{pages}{2068--2078}.
\bibitem[{{Rezvani Naraghi} et~al.(2015){Rezvani Naraghi}, Sukhov, S{\'{a}}enz,
  and Dogariu}]{RezvaniNaraghi2015}
\bibinfo{author}{R.~{Rezvani Naraghi}}, \bibinfo{author}{S.~Sukhov},
  \bibinfo{author}{J.~J. S{\'{a}}enz}, \bibinfo{author}{A.~Dogariu},
\newblock \bibinfo{title}{{Near-Field Effects in Mesoscopic Light Transport}},
\newblock \bibinfo{journal}{Phys. Rev. Lett.} \bibinfo{volume}{115}
  (\bibinfo{year}{2015}) \bibinfo{pages}{203903}.
\bibitem[{Gustavsson et~al.(2016)Gustavsson, Kristensson, and
  Wellander}]{Gustavsson2016}
\bibinfo{author}{M.~Gustavsson}, \bibinfo{author}{G.~Kristensson},
  \bibinfo{author}{N.~Wellander},
\newblock \bibinfo{title}{{Multiple scattering by a collection of randomly
  located obstacles – numerical implementation of the coherent fields}},
\newblock \bibinfo{journal}{J. Quant. Spectrosc. Radiat. Transf.}
  \bibinfo{volume}{185} (\bibinfo{year}{2016}) \bibinfo{pages}{95--100}.
\bibitem[{Ma et~al.(2017)Ma, Tan, Zhao, Wang, and Wang}]{Ma2017255}
\bibinfo{author}{L.~Ma}, \bibinfo{author}{J.~Tan}, \bibinfo{author}{J.~Zhao},
  \bibinfo{author}{F.~Wang}, \bibinfo{author}{C.~Wang},
\newblock \bibinfo{title}{{Multiple and dependent scattering by densely packed
  discrete spheres: Comparison of radiative transfer and Maxwell theory}},
\newblock \bibinfo{journal}{J. Quant. Spectrosc. Radiat. Transf.}
  \bibinfo{volume}{187} (\bibinfo{year}{2017}) \bibinfo{pages}{255--266}.
\bibitem[{Ramezan~Pour and Mackowski(2017)}]{Ramezanpour2017}
\bibinfo{author}{B.~Ramezan~Pour}, \bibinfo{author}{D.~W. Mackowski},
\newblock \bibinfo{title}{{Radiative transfer equation and direct simulation
  prediction of reflection and absorption by particle deposits}},
\newblock \bibinfo{journal}{J. Quant. Spectrosc. Radiat. Transf.}
  \bibinfo{volume}{189} (\bibinfo{year}{2017}) \bibinfo{pages}{361--368}.
\bibitem[{Mackowski and Mishchenko(2011)}]{Mackowski2011}
\bibinfo{author}{D.~Mackowski}, \bibinfo{author}{M.~Mishchenko},
\newblock \bibinfo{title}{{A multiple sphere T-matrix Fortran code for use on
  parallel computer clusters}},
\newblock \bibinfo{journal}{J. Quant. Spectrosc. Radiat. Transf.}
  \bibinfo{volume}{112} (\bibinfo{year}{2011}) \bibinfo{pages}{2182--2192}.
\bibitem[{Fuller(1991)}]{Fuller_AO_1991}
\bibinfo{author}{K.~A. Fuller},
\newblock \bibinfo{title}{{Optical resonances and two-sphere systems}},
\newblock \bibinfo{journal}{Appl. Opt.} \bibinfo{volume}{30}
  (\bibinfo{year}{1991}) \bibinfo{pages}{4716--4731}.
\bibitem[{Mackowski and Mishchenko(1996)}]{Mackowski1996}
\bibinfo{author}{D.~W. Mackowski}, \bibinfo{author}{M.~I. Mishchenko},
\newblock \bibinfo{title}{{Calculation of the T matrix and the scattering
  matrix for ensembles of spheres}},
\newblock \bibinfo{journal}{J. Opt. Soc. Am. A} \bibinfo{volume}{13}
  (\bibinfo{year}{1996}) \bibinfo{pages}{2266}.
\bibitem[{Barrett et~al.(1994)Barrett, Berry, Chan, Demmel, Donato, Dongarra,
  Eijkhout, Pozo, Romine, and Van~der Vorst}]{Barrett1994}
\bibinfo{author}{R.~Barrett}, \bibinfo{author}{M.~Berry},
  \bibinfo{author}{T.~F. Chan}, \bibinfo{author}{J.~Demmel},
  \bibinfo{author}{J.~Donato}, \bibinfo{author}{J.~Dongarra},
  \bibinfo{author}{V.~Eijkhout}, \bibinfo{author}{R.~Pozo},
  \bibinfo{author}{C.~Romine}, \bibinfo{author}{H.~Van~der Vorst},
  \bibinfo{title}{Templates for the solution of linear systems: building blocks
  for iterative methods}, \bibinfo{publisher}{SIAM}, \bibinfo{year}{1994}.
\bibitem[{Xu and Gustafson(2001)}]{Xu2001}
\bibinfo{author}{Y.~Xu}, \bibinfo{author}{B.~{\AA}. Gustafson},
\newblock \bibinfo{title}{{A generalized multiparticle Mie-solution: further
  experimental verification}},
\newblock \bibinfo{journal}{J. Quant. Spectrosc. Radiat. Transf.}
  \bibinfo{volume}{70} (\bibinfo{year}{2001}) \bibinfo{pages}{395--419}.
\bibitem[{Pellegrini et~al.(2007)Pellegrini, Mattei, Bello, and
  Mazzoldi}]{Pellegrini2007}
\bibinfo{author}{G.~Pellegrini}, \bibinfo{author}{G.~Mattei},
  \bibinfo{author}{V.~Bello}, \bibinfo{author}{P.~Mazzoldi},
\newblock \bibinfo{title}{{Interacting metal nanoparticles: Optical properties
  from nanoparticle dimers to core-satellite systems}},
\newblock \bibinfo{journal}{Mater. Sci. Eng. C} \bibinfo{volume}{27}
  (\bibinfo{year}{2007}) \bibinfo{pages}{1347--1350}.
\bibitem[{Chew et~al.(1995)Chew, Lin, and Yang}]{chew1995fft}
\bibinfo{author}{W.~C. Chew}, \bibinfo{author}{J.~H. Lin},
  \bibinfo{author}{X.~G. Yang},
\newblock \bibinfo{title}{An fft t-matrix method for 3d microwave scattering
  solutions from random discrete scatterers},
\newblock \bibinfo{journal}{Microw. Opt. Technol. Lett.} \bibinfo{volume}{9}
  (\bibinfo{year}{1995}) \bibinfo{pages}{194--196}.
\bibitem[{Gumerov and Duraiswami(2005{\natexlab{a}})}]{Gumerov2005}
\bibinfo{author}{N.~A. Gumerov}, \bibinfo{author}{R.~Duraiswami},
\newblock \bibinfo{title}{{Computation of scattering from clusters of spheres
  using the fast multipole method}},
\newblock \bibinfo{journal}{J. Acoust. Soc. Am.} \bibinfo{volume}{117}
  (\bibinfo{year}{2005}{\natexlab{a}}) \bibinfo{pages}{1744}.
\bibitem[{Gumerov and Duraiswami(2005{\natexlab{b}})}]{gumerov2005fast}
\bibinfo{author}{N.~A. Gumerov}, \bibinfo{author}{R.~Duraiswami},
  \bibinfo{title}{Fast multipole methods for the Helmholtz equation in three
  dimensions}, \bibinfo{publisher}{Elsevier},
  \bibinfo{year}{2005}{\natexlab{b}}.
\bibitem[{Gimbutas and Greengard(2013)}]{gimbutas2013fast}
\bibinfo{author}{Z.~Gimbutas}, \bibinfo{author}{L.~Greengard},
\newblock \bibinfo{title}{Fast multi-particle scattering: A hybrid solver for
  the maxwell equations in microstructured materials},
\newblock \bibinfo{journal}{J. Comp. Phys.} \bibinfo{volume}{232}
  (\bibinfo{year}{2013}) \bibinfo{pages}{22--32}.
\bibitem[{Markkanen and Yuffa(2017)}]{markkanen2017fast}
\bibinfo{author}{J.~Markkanen}, \bibinfo{author}{A.~J. Yuffa},
\newblock \bibinfo{title}{Fast superposition t-matrix solution for clusters
  with arbitrarily-shaped constituent particles},
\newblock \bibinfo{journal}{J. Quant. Spectrosc. Radiat. Transf.}
  \bibinfo{volume}{189} (\bibinfo{year}{2017}) \bibinfo{pages}{181--188}.
\bibitem[{Gumerov and Duraiswami(2008)}]{gumerov2008fast}
\bibinfo{author}{N.~A. Gumerov}, \bibinfo{author}{R.~Duraiswami},
\newblock \bibinfo{title}{Fast multipole methods on graphics processors},
\newblock \bibinfo{journal}{J. Comp. Phys.} \bibinfo{volume}{227}
  (\bibinfo{year}{2008}) \bibinfo{pages}{8290--8313}.
\bibitem[{{Doicu, A; Wriedt, T; Eremin}(2006)}]{Doicu2006}
\bibinfo{author}{Y.~A. {Doicu, A; Wriedt, T; Eremin}}, \bibinfo{title}{{Light
  Scattering by Systems of Particles}}, \bibinfo{publisher}{Springer-Verlag},
  \bibinfo{address}{Berlin, Heidelberg}, \bibinfo{year}{2006}.
\bibitem[{Mackowski(1991)}]{Mackowski1991}
\bibinfo{author}{D.~W. Mackowski},
\newblock \bibinfo{title}{{Analysis of Radiative Scattering for Multiple Sphere
  Configurations}},
\newblock \bibinfo{journal}{Proc. R. Soc. London A Math. Phys. Eng. Sci.}
  \bibinfo{volume}{433} (\bibinfo{year}{1991}).
\bibitem[{Stein(1961)}]{Stein1961}
\bibinfo{author}{S.~Stein},
\newblock \bibinfo{title}{{Addition theorems for spherical wave functions}},
\newblock \bibinfo{journal}{Q. Appl. Math.} \bibinfo{volume}{19}
  (\bibinfo{year}{1961}) \bibinfo{pages}{15--24}.
\bibitem[{Cruzan(1962)}]{Cruzan1962}
\bibinfo{author}{O.~R. Cruzan},
\newblock \bibinfo{title}{{Translational addition theorems for spherical vector
  wave functions}},
\newblock \bibinfo{journal}{Q. Appl. Math.} \bibinfo{volume}{20}
  (\bibinfo{year}{1962}) \bibinfo{pages}{33--40}.
\bibitem[{Bostr\"om et~al.(1991)Bostr\"om, Kristensson, and
  Str\"om}]{Bostrom1991}
\bibinfo{author}{A.~Bostr\"om}, \bibinfo{author}{G.~Kristensson},
  \bibinfo{author}{S.~Str\"om},
\newblock \bibinfo{title}{Transformation properties of plane, spherical and
  cylindrical scalar and vector wave functions},
\newblock in: \bibinfo{editor}{V.~Varadan}, \bibinfo{editor}{A.~Lakhtakia},
  \bibinfo{editor}{V.~Varadan} (Eds.), \bibinfo{booktitle}{Acoustic,
  Electromagnetic and Elastic Wave Scattering, Field Representations and
  Introduction to Scattering}, volume~\bibinfo{volume}{1},
  \bibinfo{publisher}{Elsevier}, \bibinfo{year}{1991}, pp.
  \bibinfo{pages}{165--210}.
\bibitem[{Mishchenko et~al.(2002)Mishchenko, Travis, and
  Lacis}]{Mishchenko2002}
\bibinfo{author}{M.~I. Mishchenko}, \bibinfo{author}{L.~D. Travis},
  \bibinfo{author}{A.~A. Lacis}, \bibinfo{title}{{Scattering, Absorption, and
  Emission of Light by Small Particles}}, \bibinfo{publisher}{Cambridge
  University Press}, \bibinfo{year}{2002}.
\bibitem[{Bohren and Huffman(1983)}]{Bohren1983}
\bibinfo{author}{C.~F. Bohren}, \bibinfo{author}{D.~R. Huffman},
  \bibinfo{title}{{Absorption and scattering of light by small particles}},
  {Wiley science paperback series}, \bibinfo{publisher}{Wiley},
  \bibinfo{year}{1983}.
\bibitem[{Novotny and Hecht(2006)}]{Novotny2006}
\bibinfo{author}{L.~Novotny}, \bibinfo{author}{B.~Hecht},
  \bibinfo{title}{{Principles of Nano-Optics}}, volume~\bibinfo{volume}{1},
  \bibinfo{publisher}{Cambridge University Press},
  \bibinfo{address}{Cambridge}, \bibinfo{year}{2006}.
\bibitem[{Stratton(1941)}]{Stratton1941}
\bibinfo{author}{J.~A. Stratton}, \bibinfo{title}{{Electromagnetic Theory}},
  \bibinfo{publisher}{McGraw-Hill}, \bibinfo{address}{New York},
  \bibinfo{year}{1941}.

\end{thebibliography}

\end{document}